\documentclass[aps,prc,twocolumn,showpacs,amsmath,amssymb,nofootinbib]{revtex4-2}
\usepackage{graphicx}
\usepackage{color}
\usepackage{times}
\usepackage{inputenc}
\usepackage{bm}
\usepackage{ulem}
\usepackage{multirow}
\usepackage{float}
\usepackage{url}
\usepackage{natbib}
\usepackage{mathrsfs}
\usepackage{physics}
\usepackage{comment}
\usepackage[table,xcdraw]{xcolor}
\usepackage{booktabs,makecell}
\usepackage{comment}
\usepackage[colorlinks=true,citecolor=blue,urlcolor=blue,linkcolor=blue]{hyperref}
\begin{document}
%%%%%%%%%%%%%%%%
\title{Radial oscillations of dark matter admixed neutron stars}
%%%%%%%%%%%%%%%%
\author{Pinku Routaray$^{1}$}
\author{H. C. Das$^{2,3}$}
\author{Souhardya Sen$^1$}
\author{Bharat Kumar$^{1}$}
\email{kumarbh@nitrkl.ac.in }
\author{Grigoris Panotopoulos$^{4}$}
\author{Tianqi Zhao$^{5}$}
%%%%%%%%%%%%%%%%%%%%%%%%%%%%
\affiliation{\it $^{1}$Department of Physics \& Astronomy, National Institute of Technology, Rourkela 769008, India}
\affiliation{\it $^{2}$Institute of Physics, Sachivalaya Marg, Bhubaneswar 751005, India}
\affiliation{\it $^{3}$Homi Bhabha National Institute, Training School Complex, Anushakti Nagar, Mumbai 400094, India}
\affiliation{\it $^{4}$Departamento de Ciencias F{\'i}sicas, Universidad de la Frontera, Casilla 54-D, 4811186 Temuco, Chile}
\affiliation{\it $^{5}$Department of Physics and Astronomy, Ohio University, Athens, Ohio~45701, USA}
%%%%%%%%%%%%%%%%
\date{\today}
%%%%%%%%%%%%%%%%
\begin{abstract}
Within the relativistic mean-field model, we investigate the properties of dark matter (DM) admixed neutron stars, considering nonrotating objects made of isotropic matter. We adopt the IOPB-I hadronic equation of state (EOS) by assuming that the fermionic DM within supersymmetric models has already been accreted inside the neutron star (NS). The impact of DM on the mass-radius relationships and the radial oscillations of pulsating DM admixed neutron stars (with and without the crust) are explored. It is observed that the presence of DM softens the EOS, which in turn lowers the maximum mass and its corresponding radius. Moreover, adding DM results in higher frequencies of pulsating objects, and hence we show the linearity of the fundamental mode frequency of canonical NS with DM Fermi momentum. We also investigate the profile of eigenfunctions solving the Sturm-Liouville boundary value problem and verify its validity. Further, we study the stability of NSs considering the fundamental mode frequency variation with the mass of the star and verify the stability criterion $\partial M/\partial\rho_c > 0$. Finally, the effect of the crust on the large frequency separation for different DM Fermi momenta is shown as well.
\end{abstract}
%%%%%%%%%%%%%%%
\maketitle
%%%%%%%%%%%%%%%
%%%%%%%%%%%%%%%%%%%%%%
\section{Introduction}
\label{intro}
%%%%%%%%%%%%%%%%%%%%%%
The neutron star (NS) is one of the mysterious stellar leftovers, having an enormously dense core and a robust crust. The coalescence of binary NS (BNS) merger events produce gravitational waves (GWs) that encode sufficient knowledge to place substantial restrictions on the equation of state (EOS) and the internal compositions of NSs \cite{Abbott_prl2017, Abbott_prl2018, Bauswein_2017, Annala_2018, Fattoyev_2018, Malik_2018, Elias_2018, Tews_2018, Nandi_2019, Capano_2020, Radice_2018}. In the future, terrestrial detectors, such as LIGO/Virgo/KAGRA, could be able to observe more BNS merger events, which could more precisely restrict the properties of compact stars. In addition to that, oscillating NSs also emit GWs with various mode frequencies can be used to explore the internal compositions as well as the various properties of the star \cite{Andersson_1996, Andersson_1998}.

Oscillating NSs emit different modes frequencies, such as $f$, $p$, $g$, etc., depending on the restoring force, after their formation in the supernovae. There are various processes, such as dynamical instability, mass accretion, magnetic configuration, and fractures in the crust, that may be the different sources of oscillations \cite{Chirenti_2017, Hinderer_2016, Franco_2000, Tsang_2012}. Oscillations are mainly categorized into two types, namely radial and nonradial. In this study, we propose to explore various properties of radial oscillations of NSs. Several works have already been published on the exploration of different properties of radially oscillating NS \cite{Chandrasekhar_1964, Chanmugam_1977, Vaeth_1992, Gondek_1997, Gondek_1999, Grigorious_2017, Sagun_2020, Vaeth_1992, Kokkotas_2001}. Within the framework of general relativity, radial oscillations have been investigated as the simplest mode of NSs \cite{Chandrasekhar_1964, Chanmugam_1977,Vaeth_1992, Gondek_1997, Gondek_1999, Grigorious_2017, Sagun_2020}. Later, other numerical techniques incorporated zero temperature EOS to correct their numerical results \cite{Vaeth_1992, Kokkotas_2001}. According to their findings, oscillations become unstable once the NS reaches its maximum mass at the corresponding central density. The detection of radial oscillations is quite complex, as they cannot generate GWs on their own. They are coupled with nonradial oscillations, making GWs stronger and making it more likely that they can be detected \cite{Passamonti_2006, Passamonti_2007}. However, in Ref. \cite{Chirenti_2019}, it has been observed that in the postmerger event of BNS, a hypermassive NS is formed along with the emission of a short gamma-ray burst (SGRB), which may be impacted by radial oscillations. Therefore, in this study, we want to explore more with the inclusion of dark matter (DM) inside the oscillating NS since DM effects produce an extra peak in the postmerger spectrum \cite{Ellis_plb2018}. 

Evidence for DM in galaxies may be inferred from a variety of data, including galaxies' rotation curves, velocity dispersions, galaxy clusters, gravitational lensing, the cosmic microwave background, etc. According to the findings of cosmological observations, the unseen matter cannot be composed of baryons; instead, it needs to be a new type of matter that only has a very weak interaction with the other particles. However, substantial research on DM models introduced and analyzed by particle physicists has led to the establishment of stringent limits on the coupling constant as well as the mass of the DM particle. The weakly interacting massive particle (WIMP) scenario has gained popularity among DM models since it is the most abundant DM candidate and the thermal relic of the Universe. Therefore, in our model, we choose nonannihilating WIMPs (neutralino) as the DM candidate, which is already accreted inside the NS \cite{arpan_2019, Grigorious_2017, harishmnras_2020, harishprd_2021, harishmnras_2021, harishjcap_2021}. From various observational data, one can put constraints on the amount of DM inside the NS \cite{harishmnras_2020, harishprd_2021}. Alternatively, asteroseismology is a widely used technique to probe the inner structure of stars. By studying oscillations of pulsating objects and computing the frequencies of their modes, we could learn more about the EOS of interacting matter and internal composition since the numerical values of the frequency modes are extremely sensitive to the underlying physics.

To study NS properties, one needs the EOS, which dictates the relationship between energy density and pressure. Here, we take the extended relativistic mean field (E-RMF) model, as explained in detail in Refs. \cite{KUMAR_2017, kumar_2018}. However, in this study, we consider the IOPB-I EOS \cite{kumar_2018} to calculate various properties of NSs. In the case of DM, we use the method as discussed in Sec. \ref{DM}. The final EOS is the addition of nucleons and DM. With that EOS, we investigate the effects of DM on $f$ and $p$-mode frequencies. Assuming nonrotating NSs, we solve the Sturm-Liouville eigenvalue problem \cite{Ince_1956, Byron_1992, Cox_1980}. Our model is predicated on the assumption that oscillations are characterized by a small magnitude, which allows for the application of linear perturbation theory\cite{Bardeen_1966, Haensel_1989}. Our work is organized as follows: we focus on the formalism to obtain the EOS for DM admixed NS (DMANS) in Sec. \ref{DM}. We calculate the mass-radius relationships for a static, isotropic, and nonrotating star, as discussed in Sec. \ref{TOV}. The methodology for radial oscillation is enumerated in Sec. \ref{RO}. We discuss our numerical results in Sec. \ref{RD}, and finally, we summarize our work in Sec. \ref{summary}.
%%%%%%%%%%%%%%%%%%%
\section{Formalism}
%%%%%%%%%%%%%%%%%%%
%%%%%%%%%%%%%%%%%%%%%%%%%%%%%%%%%%%%%%%%%%%%%%%%%%%%
\subsection{Model for DM admixed NS}
\label{DM}
%%%%%%%%%%%%%%%%%%%%%%%%%%%%%%%%%%%%%%%%%%%%%%%%%%%%
Compact objects, such as NS, capture a finite amount of DM in their evolution stage. After accretion, the DM particle loses energy when interacting with neutrons because of the high baryon density. The NS's immense gravitational field traps the DM after losing some energy \cite{Kouvaris_2011, Goldman_1989, arpan_2019}. Since WIMPs are the most prevalent DM particle and the thermal remnant, we choose nonannihilating WIMPs (neutralino) as DM candidates. Other phenomena enhance the density of DM inside the NS, including converting neutrons to scalar DM and generating scalar DM via bremsstrahlung \cite{Ellis_prd2018, Ellis_plb2018,arpan_2019}. The history of NS's formation and its habitat's surroundings affect how much DM is present.

The DM particle interacts with baryons by exchanging SM Higgs. The form of the interacting Lagrangian is given by \cite{arpan_2019,Grigorious_2017,harishmnras_2020,harishjcap_2021,harishmnras_2021,harishprd_2021,Quddus_2020},
%%%%%%%%%%%%%%%%
\begin{eqnarray}
{\cal{L}}_{\rm DM} & = & \bar \chi \left[ i \gamma^\mu \partial_\mu - M_\chi + y h \right] \chi +  \frac{1}{2}\partial_\mu h \partial^\mu h  \nonumber\\
& &
- \frac{1}{2} M_h^2 h^2 + f \frac{M_{\rm nucl.}}{v} \bar \varphi h \varphi \, , 
\label{eq:LDM}
\end{eqnarray}
%%%%%%%%%%%%%%
where $\varphi$ and $\chi$ are the nucleonic and DM wave functions, respectively. $h$ is the Higgs field. The masses $M_\chi$ and $M_h$ are the neutralino mass and Higgs mass taken as 200 GeV and 125 GeV, respectively. The coupling constants between the DM and SM Higgs is $y$, which can be obtained in the large Higgs mixing angle limit. The various gauge coupling constants are present in the electroweak sector of the standard model \cite{MARTIN_1998} because the neutralino is a supersymmetric particle. The values of $y$ are given in the range between 0.001 to 0.1, depending on the various parameters \cite{Grigorious_2017,harishmnras_2020,arpan_2019}. Therefore, we use $y = 0.07$ in our computations. $f M_{\rm nucl.}/v$ is the effective Yukawa coupling between the Higgs field and nucleons, where $f$ is the proton-Higgs form factor. Its value can be assumed to be approximately 0.35 \cite{Cline_2013}, and $v$ is the vacuum expectation of Higgs taken as 246 GeV \cite{harishmnras_2020, harishjcap_2021}. 

The DMANS is taking into account as single fluid based on the speculation that the DM is sufficiently coupled to the baryonic matter to allow for single fluid approximation. This approximation originates from the fact that the DM considered as collisionless and weak interactions  so that it can be modeled as pressureless fluid. It has been shown that the interaction between DM and baryonic matter can result in a efficient thermalization process, whereby DM can effectively transfer energy and momentum with the baryonic fluid.
\\
It is important to demonstrate that dark matter coupled to baryonic matter strong enough that they may be considered a single fluid. In order to provide support for the single-fluid approximation, the typical timescales for interactions between the two fluids might be compared to achieve this goal. Dark matter and baryonic matter may be considered as a single fluid if the timescale for their interactions is substantially shorter than that of the neutron star's oscillations. Order of magnitude estimation shows the relaxation time is $\tau\approx \frac{1}{\sigma_{SI} nc}\approx2$ ms, where $\sigma_{SI}$ is spin-independent cross section, $n$ is the density and c is the velocity of light . Due to the high density $2\times10^{14}$ g/cm$^3$ and nonzero cross section $10^{-46}$ cm$^{2}$, the relaxation time is comparable to the millisecond period of NS oscillation. This possibility is not completely ruled out because of the finite cross section.
\\
However, it is crucial to keep in mind that this assumption could not always be accurate, particularly for some certain DM models.  Specifically, the models where  DM and baryonic matter exhibits weak interaction among themselves and only interact gravitationally where the relaxation time may exceed the period of millisecond oscillation to a major degree. A two-fluid approximation could be regarded more suitable in such an instance. We acknowledge that the justification of the single-fluid  assumption must be established for each distinct model of dark matter. In addition, supplementary references have been provided to validate the authenticity of the single-fluid approximation for various dark matter models. For more details, please see \cite{Bertoni2013, Boddy2019, Goldman_1989, Kouvaris2008, Kavanagh2017,Grigorious_2017}.
%%%%%%%%%%%%%%%%%%%%%%
\subsubsection{Experimental evidences}
%%%%%%%%%%%%%%%%%%%%%%
The DM-Higgs coupling factor ($y$) plays a significant role in this model. One can constrain its magnitude from different DM detection experiments.  Although the direct detection experiment has not reported any events to date, however, they provided some upper limits on the WIMP-nucleon scattering cross section. Since this model is based on the assumption that, the DM particles interact with nucleons by exchanging the standard model Higgs. Hence,  the Higgs exchange causes an elastic scattering between the WIMPs and the nuclei, mainly at the quark level. In light of this, the interaction Lagrangian that includes both the DM wave function ($\chi$) and the quark wave function ($q$) can be expressed as follows \cite{Bhat_2020},
%%%%%%%%%%%%%%%%
\begin{equation}
{\cal{L}}_{\rm int}=\alpha_q \Bar{\chi}\chi\Bar{q}q,
\end{equation}
%%%%%%%%%%%%%%%
where $\alpha_q=\frac{y f m_q}{vM_h^2}$. $f$ is the nucleon-Higgs form factor, and $m_q$ is the mass of the quark. The values of $y$ and $f$ are taken as 0.07 and 0.35, respectively, in this study.

For the fermionic DM, one can write the spin-independent cross section as follows \cite{Bhat_2020},
 %%%%%%%%%%%%%%%%
\begin{equation}
\sigma_{\rm SI}=\frac{y^2f^2M_n^2}{4\pi}\frac{\mu_r^2}{v^2M_h^4} ,
\end{equation}
%%%%%%%%%%%%%%%
where $M_n$ (= 939 MeV) is the nucleon mass and $\mu_r$ is the reduced mass $\frac{M_nM_\chi}{M_n+M_\chi}$, $M_\chi$ is the mass of the DM particle. We took $y$ = 0.07 and DM mass as 200 GeV and computed their corresponding cross section, which is found to be $9.70\times 10^{-46}$ cm$^2$, which indicates that the $\sigma_{\rm SI}$ is agreed with direct detection experiments such as XENON-1T \cite{Xenon1T_2016}, PandaX-II \cite{PandaX_2016}, and LUX \cite{LUX_2017} with  90\% confidence level. The large hadron collider (LHC) also put a limit on the WIMP-nucleon scattering cross section between $10^{-40}$ to $10^{-50}$ cm$^2$ \cite{Djouadi_2012}. Hence, our model also satisfies the LHC limit. Moreover, it is observed that the small change in the magnitude of $y$ has no impact on the frequency of the radial oscillation.
%%%%%%%%%%%%%%%%%%%%%%%%%%%%%%%%%%%%%%%%
\subsection{Equation of state for DMANS}
%%%%%%%%%%%%%%%%%%%%%%%%%%%%%%%%%%%%%%%%
In this study, we use the extended relativistic mean field (E-RMF) model EOS of the NS. Several works have already used the E-RMF model and found that almost NS properties, such as mass, radius, tidal deformability, the moment of inertia, etc., are well reproduced and consistent with different observational data \cite{kumar_2018, harishmnras_2020, harishmnras_2021, harishjcap_2021}. The details calculations and the applications of E-RMF to NS can be found in \cite{kumar_2018}. However, the energy density and pressure for DM can be calculated by using the Eq. (\ref{eq:LDM}) given as \cite{harishmnras_2020, harishmnras_2021, Grigorious_2017},
%%%%%%%%%%%%%%%%
\begin{eqnarray}
{\cal{E}}_{\rm DM} = \frac{2}{(2\pi)^{3}}\int_0^{k_f^{\rm DM}} d^{3}k \sqrt{k^2 + (M_\chi^\star)^2 } + \frac{1}{2}M_h^2 h_0^2 \, ,
\label{eq:edm}
\end{eqnarray}
%%%%%%%%%%%%%%%%
\begin{eqnarray}
P_{\rm DM} = \frac{2}{3(2\pi)^{3}}\int_0^{k_f^{\rm DM}} \frac{d^{3}k \hspace{1mm}k^2} {\sqrt{k^2 + (M_\chi^\star)^2}} - \frac{1}{2}M_h^2 h_0^2 \, ,
\label{eq:pres}
\end{eqnarray} 
%%%%%%%%%%%%%%
where $k_f^{\rm DM}$ is the DM Fermi momentum, and $M_\chi^* (=M_\chi-yh_0$) is the DM effective mass. We assume that the DM density is $\sim 1000$ times less than the baryon density inside NS \cite{arpan_2019, Grigorious_2017}. From this assumption, one can calculate the DM density. Hence, in our computations, we vary the value of $k_f^{\rm DM}$ from 0-0.05 GeV.

Therefore, for DMANS, the total energy density and pressure can be written as \cite{harishmnras_2020, harishjcap_2021}
%%%%%%%%%%%%%%%%
\begin{eqnarray}
{\cal{E}}={\cal{E}}_{\rm NS}+ {\cal{E}}_{\rm DM} \, ,
\nonumber
\\
{\rm and}
\hspace{1cm}
P=P_{\rm NS} + P_{\rm DM} \, ,
\label{eq:EOS_total}
\end{eqnarray}
%%%%%%%%%%%%%%
where ${\cal{E}}={\cal{E}}_{\rm NS}$, and $P = P_{\rm NS}$ for NS with only nucleons can be found in \cite{harishmnras_2020, harishmnras_2021}.

\begin{figure}
    \centering
    \includegraphics[scale=0.6]{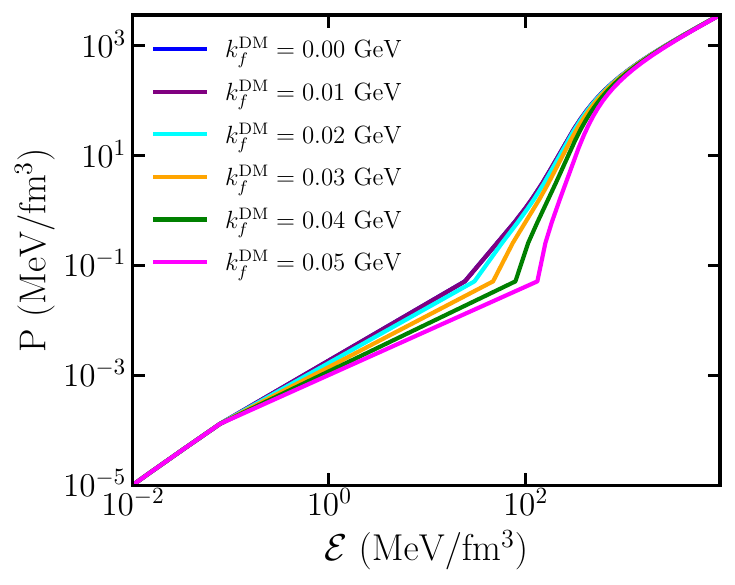}
    \caption{Unified DM admixed EOSs for various Fermi momenta. The kink at P $\approx$ $10^{-1}$ MeV/fm$^3$ shows} the crust-core transitions.
    \label{fig:EOS}
\end{figure}

Figure \ref{fig:EOS} depicts the variation of pressure and energy density by varying DM Fermi momenta. For the lower-density region, we use the SLY4 crust to make the unified EOSs \cite{Douchin2001}, which can be used to study the frequencies of the radially oscillating DM admixed NS. It is observed that the EOS becomes softer with the addition of DM. The softening of the EOSs depends on the amount of DM inside the NS.

Due to the fact that all systems strive to save energy, this is the case. The Fermi momenta ($k_f$), also known as the Fermi energy, becomes larger as the density rises. Since nucleons are fermions, they must be assigned a higher orbit as the nuclear density rises. The density is known to increase as the cube of the Fermi momenta. The nucleon's total energy, defined as $E=\sqrt{k_f^2 + M^2 }$ grows as the density of the nucleus rises. The nucleons decay into these particles when their energy level is greater than that of the DM. Therefore, even though DM is heavier than nucleons, it is more energy efficient for the system to have the DM in the lower-energy states instead of nucleons at greater Fermi energy and at greater density. According to the density of the system, DM may substitute in place of the nucleons. So, some of the gravitational mass is turned into kinetic energy, which makes the total mass smaller.
%%%%%%%%%%%%%%
%%%%%%%%%%%%
%%%%%%%%%%%%%%%%%%%%%%%%%%%%%%%%%%%%%%%%%%%%%%%
\subsection{Hydrostatic equilibrium structure}
\label{TOV}
%%%%%%%%%%%%%%%%%%%%%%%%%%%%%%%%%%%%%%%%%%%%%%%
In equilibrium, the metric tensor for a static and spherically symmetric star is given by \cite{Schwarzschild_1916}
%%%%%%%%%%%%%%%%
\begin{eqnarray}
ds^2 = -e^{2\nu} c^2 dt^2 + e^{2\lambda} dr^2 + r^2(d\theta^2 + \text{sin}^2\theta d\phi^2) \, ,
\label{eq:spherical_metric}
\end{eqnarray}
%%%%%%%%%%%%%%%
where $e^{2\nu}$ and $e^{2\lambda}$ are the metric functions. 

To describe the hydrostatic equilibrium of NSs, Einstein's field equations in Schwarzschild-like coordinates imply the Tolman-Oppenheimer-Volkoff (TOV) equations, which are given by \cite{Oppenheimer_1939, Tolman_1939}
%%%%%%%%%%%%%%%%
\begin{align}
\frac{dP}{dr} &= - \frac{Gm}{c^2r^2} \frac{\left( P+\mathcal{E} \right) \left( 1 + \frac{4\pi r^3P}{mc^2}\right)}{\left(1-\frac{2Gm}{c^2r}\right)} \, ,
\nonumber \\
\frac{d\nu}{dr}&=-\frac{1}{P+\mathcal{E}}\frac{dP}{dr} \, ,
\nonumber \\
\frac{dm}{dr} &= \frac{4\pi r^2 \mathcal{E}}{c^2} \, ,
\label{eq:TOV}
\end{align}
%%%%%%%%%%%%%%
and the corresponding metric functions at the surface, i.e., at $r=R$
%%%%%%%%%%%%%%%%
\begin{eqnarray}
e^{2\nu(R)} = e^{-2\lambda(R)} = \left(1-\frac{2GM}{c^2R}\right) \, .
\label{eq:potential_surface}
\end{eqnarray}
%%%%%%%%%%%%%%%%

By using the initial conditions $m(r=0)=0$ and $P(r=0) = P_c$, where $P_c$ is the central pressure, here TOV equations can be solved for DMANS EOSs, and the integration will continue up to surface boundary where $m(r=R)=M$ and $P(r=R)=0$. 

\begin{figure}
    \centering
    \includegraphics[scale=0.65]{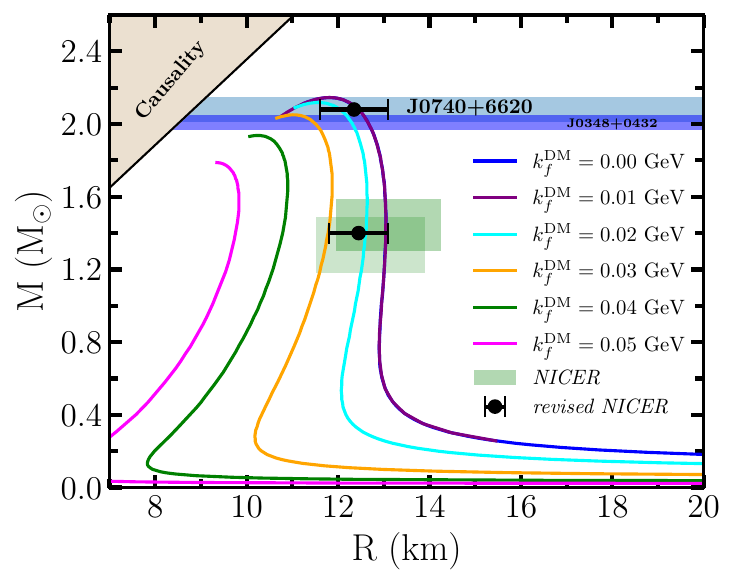}
    \caption{Mass-radius relations for DM admixed NS for IOPB-I EOS with varying DM Fermi momentum. Different horizontal color bands for various pulsars' observational data and the simultaneous observations of mass-radius given by NICER are also shown.}
    \label{fig:mr}
\end{figure}

We plot the mass-radius relations for DM admixed NS by solving the TOV equations for various Fermi momentum of DM in Fig. \ref{fig:mr}. 
%%%%%%%%%%%%%%
%%%%%%%%%%%%
The magnitude of the maximum mass and its corresponding radius decreases with increasing DM momenta. For $\sim k_f^{\rm DM} = 0.03$ GeV, the $M-R$ curves reproduce the observational data well. The curves do not satisfy any of the observational data with more percentages. Therefore, from the observational data, one can fix the amount of DM inside the NS. 
%%%%%%%%%%%%%%%%%%%%%%%%%%%%%%%%%%%%%%%
\subsection{Radial oscillations of NSs}
\label{RO}
%%%%%%%%%%%%%%%%%%%%%%%%%%%%%%%%%%%%%%%
To calculate the properties of the radially oscillating NS, we follow the methodology given by Kokkotas and Ruoff \cite{Kokkotas_2001}.In order to analyze the equations driving radial oscillations of NSs, we define $\delta r(r,t)$ to be the time-dependent radial displacement of a fluid element as
%%%%%%%%%%%%%%%%
\begin{eqnarray}
\delta r(r,t) = X(r)e^{i\omega t},
\label{eq:standing}
\end{eqnarray}
%%%%%%%%%%%%%%%%
where $X(r)$ is the amplitude and $\omega$ is the circular frequency of the standing wave solution.The linearized perturbation equations can be expressed as a second-order homogeneous differential equation with the assumption of adiabatic oscillations as
%%%%%%%%%%%%%%%%
\begin{equation}\label{eq:radial_equation}
\begin{aligned}
&c_{s}^{2} X^{\prime \prime}+\left(\left(c_{s}^{2}\right)^{\prime}-Z+ \frac{4 \pi G}{c^4} r \gamma P e^{2 \lambda}-\nu^{\prime} c^2 \right) X^{\prime} \\
&+{\left[2\left(\nu^{\prime}\right)^{2}c^2+\frac{2 G m}{r^{3}} e^{2 \lambda}-Z^{\prime}-\frac{4 \pi G }{c^4} (P+\mathcal{E}) Z r e^{2 \lambda}\right]X}\\
&+ \omega^{2} e^{2 \lambda-2 \nu} X =0,
\end{aligned}
\end{equation}
%%%%%%%%%%%%%%
where $c_s^2$ is the sound of speed squared and $\gamma$ is the adiabatic index, of the forms
%%%%%%%%%%%%%%%%
\begin{eqnarray}
c_s^2 = \frac{dP}{d\cal{E}}c^2
\hspace{1cm}
{\rm and}
\hspace{1cm}
\gamma=\left(1+\frac{\mathcal{E}}{P}\right)c_{s}^{2}.
\label{eq:adiabatic_index}
\end{eqnarray}
%%%%%%%%%%%%%%
Also,
%%%%%%%%%%%%%%%%
\begin{eqnarray}
Z(r)=\left(\nu^{\prime}-\frac{2}{r}\right)c_{s}^{2}.
\label{eq:Zerili_function}
\end{eqnarray}
%%%%%%%%%%%%%%%
The oscillation equations should be such that there is no displacement at the center, i.e.,
%%%%%%%%%%%%%%%%
\begin{eqnarray}
\delta r(r=0)=0
\label{eq:delbcr}
\end{eqnarray}
%%%%%%%%%%%%%%%
and the Lagrangian perturbation of pressure should vanish at the surface,
%%%%%%%%%%%%%%%%
\begin{eqnarray}
\Delta P(r=R)= 0.
\label{eq:delbcP}
\end{eqnarray}
%%%%%%%%%%%%%%%
Owing to these boundary conditions, the displacement function can be redefined as,
%%%%%%%%%%%%%%%%
\begin{eqnarray}
\zeta = r^2e^{-\nu}X \, .
\label{eq:perturbed_equation}
\end{eqnarray}
%%%%%%%%%%%%%%%%
Using this new variable, Eq.  (\ref{eq:radial_equation}) can be rewritten as a Sturm-Liouville differential equation which has a self-adjoint nature \cite{Kokkotas_2001}
%%%%%%%%%%%%%%%%
\begin{eqnarray}
\frac{d}{dr}\left( H\frac{d \zeta}{d r}\right)+\left(\omega^{2} W+Q\right) \zeta=0 \, ,
\label{eq:self_adjoint_eq}
\end{eqnarray}
%%%%%%%%%%%%%%
where
%%%%%%%%%%%%%%%%
\begin{align}
&r^2H=(P+\mathcal{E}) e^{\lambda+3 \nu} c_{s}^{2} \, ,
\nonumber \\
&r^2W=(P+\mathcal{E}) e^{3 \lambda+\nu} \, ,
\nonumber \\
&r^2Q=(P+\mathcal{E}) e^{\lambda+3 \nu}\left((\nu^{\prime})^{2}+\frac{4}{r} \nu^{\prime}- \frac{8 \pi G}{c^4} e^{2 \lambda} P\right).
\label{eq:HWQ}
\end{align}
%%%%%%%%%%%%%%
Equation (\ref{eq:self_adjoint_eq}) is the master equation for radial oscillations such that $\Delta P$ takes a simple form
%%%%%%%%%%%%%%%%
\begin{eqnarray}
    \Delta P = -r^{-2} e^{\nu}(P+\mathcal{E}) c_{s}^{2}\zeta^{\prime}.
    \label{eq:pres_variation}
\end{eqnarray}
%%%%%%%%%%%%%%%
Moreover, since Eq.  (\ref{eq:self_adjoint_eq}) takes the Sturm-Liouville form, where $\zeta_n$ has $n$ nodes between the surface and the center with discrete eigenvalues $\omega_n^2$. The eigenvalues follow
%%%%%%%%%%%%%%%%
\begin{eqnarray}
    \omega_{0}^{2} < \omega_{1}^{2} <... <  \omega_{n}^{2} < ... .
    \nonumber
    \label{eq:disc}
\end{eqnarray}
%%%%%%%%%%%%%%%%
The standing wave solution, Eq.  (\ref{eq:standing}) also suggests that oscillations will be harmonic and stable for real $\omega$ but the star will become unstable with an imaginary frequency of the node. Additionally, because the eigenvalues are arranged in the manner described above, it is crucial to know the fundamental $f$-mode frequency ($n=0$) in order to determine the stability of the star. $\omega_0$ becomes imaginary for a central density greater than the critical density ($\rho_{\rm crit}$), which corresponds to the density at which NS attains its maximum mass. Above $\rho_{\rm crit}$, the amplitude of oscillations becomes exponential, and the star cannot return to its original configuration, finally collapsing into a black hole.

The Eq. (\ref{eq:self_adjoint_eq}) is split into two first-order coupled linear differential equations for numerical integration. To do this, we create a new variable called $\eta$, where
%%%%%%%%%%%%%%%%
\begin{eqnarray}
    \eta = H\zeta^{\prime}.
    \label{eq:eta}
\end{eqnarray}
The coupled differential equation thus becomes \cite{Kokkotas_2001}
%%%%%%%%%%%%%%%%
\begin{eqnarray}
    \frac{d \zeta}{d r} = \frac{\eta}{H} ,
    \label{eq:zeta}
\end{eqnarray}
%%%%%%%%%%%%%%%%
%%%%%%%%%%%%%%%%
\begin{eqnarray}
    \frac{d \eta}{d r} = -\left(\omega^{2} W+Q\right) \zeta .
    \label{eq:deta}
\end{eqnarray}
%%%%%%%%%%%%%%%%
Using Taylor expansion on $\zeta$ near the origin and Eq.  (\ref{eq:zeta}), we find that $\eta_0 = 3\zeta_0H_0$ with
%%%%%%%%%%%%%%%%
\begin{eqnarray}
    H_0=(P(0)+\mathcal{E}(0)) e^{\lambda(0)+3 \nu(0)} c_{s}^{2}(0)
    \label{eq:eta0}
\end{eqnarray}
%%%%%%%%%%%%%%%%
where $\eta_0$ and $\zeta_0$ are their corresponding values at the center of the star.

By choosing $\eta_0 = 1$, we get $\zeta_0 = 1 / (3H_0)$ as the initial value to start our numerical integration using the shooting method. The values of $\omega$ that satisfy $\eta(r=R)=0$ will give us the required radial oscillation modes.

%%%%%%%%%%%%%%%
\begin{figure*}
    \centering
    \includegraphics[scale=0.53]{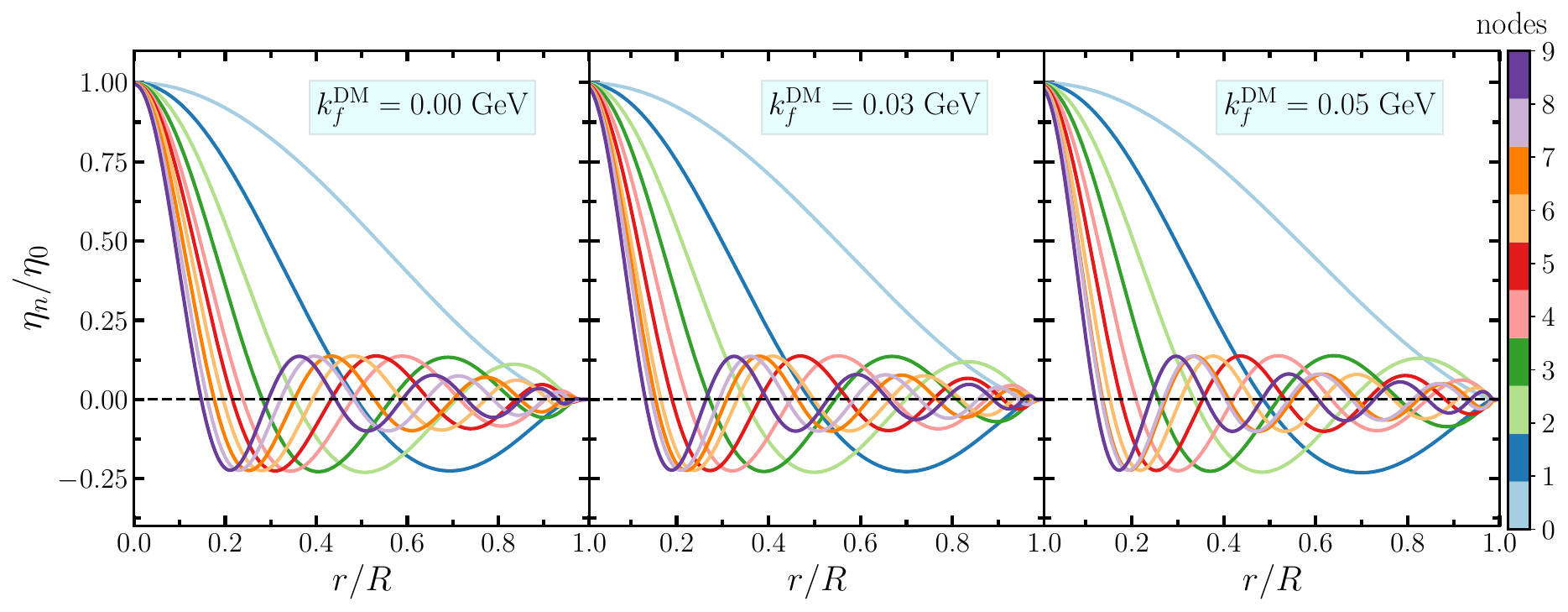}
    \\ 
    \includegraphics[scale=0.53]{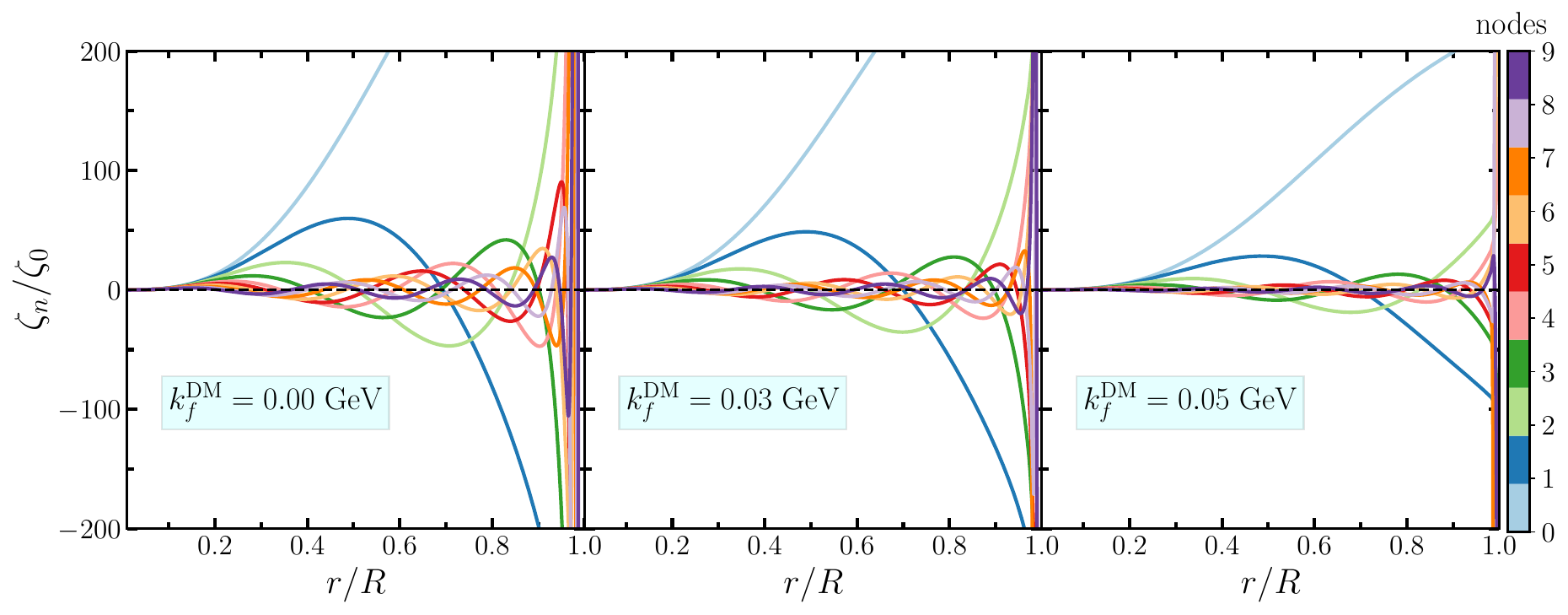}
    \caption{{\it Top:} the variation of $\eta/\eta_0$ as a function of dimensionless quantity $r/R$ for $f$-mode and higher $p$-modes ($n=1-9$) in DM admixed \textcolor{blue}{EOSs with varying Fermi momentum}. The color bar represents the order number $n$ corresponding to different modes. {\it Bottom:} same as the top panel but for $\zeta/\zeta_0$.}
    \label{fig:eta_zeta}
\end{figure*}
%%%%%%%%%%%%%%

%%%%%%%%%%%%%%%
\begin{figure*}
    \centering
    \includegraphics[scale=0.53]{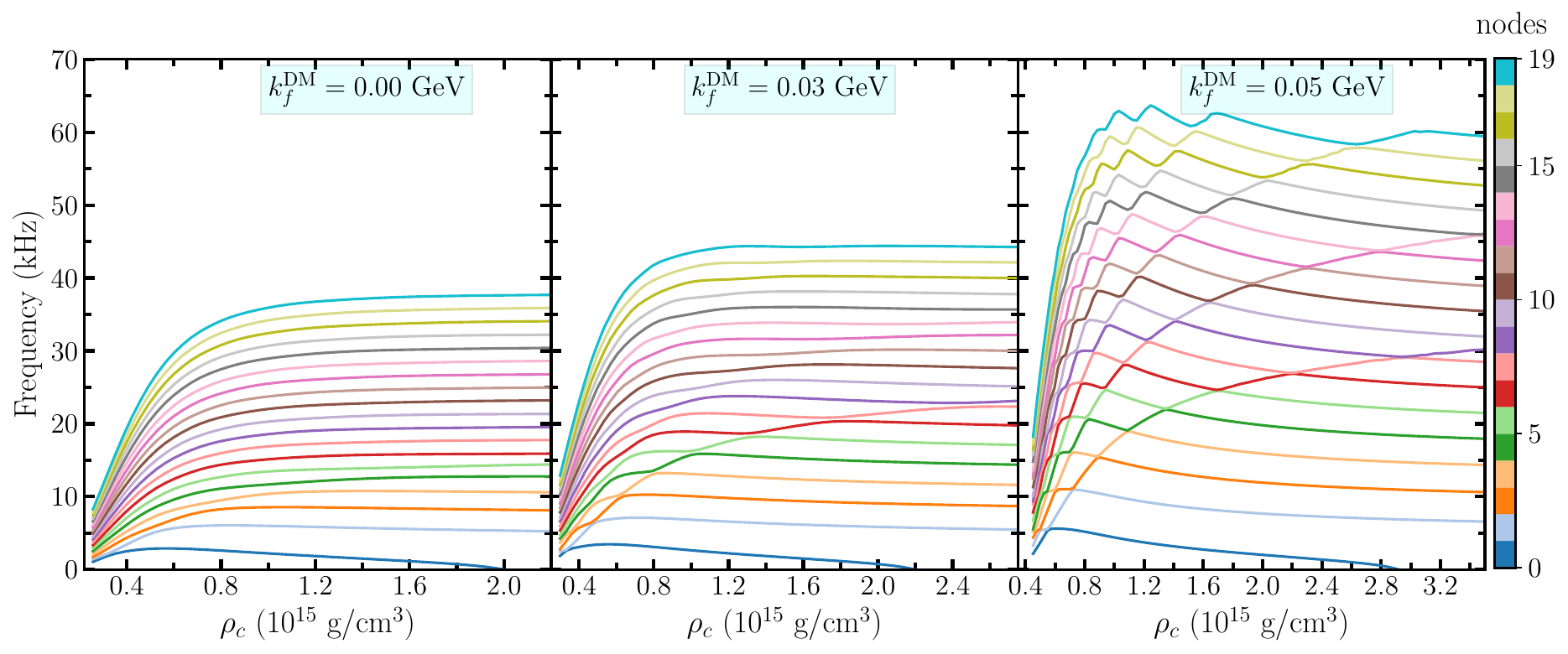}
    \caption{Frequencies of the radially oscillating NS as a function of central energy densities. Three panels represent different DM momenta with $k_f^{\rm DM}=0.00, 0.03$, and 0.05 GeV respectively. The color bar represents the order number $n$ corresponding to different modes. }
    \label{fig:freq}
\end{figure*}
%%%%%%%%%%%%%
%%%%%%%%%%%%%%
\begin{table}
\caption{20 lowest order radial oscillation frequencies (in kHz) for three DM momenta each calculated at 1.4 M$_\odot$.}
\begin{tabular*}{\linewidth}{l@{\extracolsep{\fill}}ccc}
\hline
\hline
\multirow{2}{*}{Nodes} & \multicolumn{3}{c}{ $k_f^{\rm DM}$}  \\
\cline{2-4}
                       & 0.00 GeV & 0.03 GeV & 0.05 GeV       \\
\hline
0   &2.8450    &3.1841     &3.7404\\
1   &5.8594    &7.1023     &9.3196\\
2   &7.4079    &10.2763    &13.9863\\
3   &8.6053    &12.7038    &18.4640\\
4   &10.1398   &13.4110    &20.4151\\
5   &11.6405   &16.0817    &22.8780\\
6   &13.1050   &18.1414    &27.2547\\
7   &14.6138   &19.4984    &30.8661\\
8   &16.1008   &21.4467    &31.6181\\
9   &17.6315   &23.5088    &35.8852\\
10  &19.1166   &25.4284    &40.0455\\
11  &20.6377   &27.2522    &41.3531\\
12  &22.1461   &29.1611    &44.1685\\
13  &23.6751   &31.2952    &48.1325\\
14  &25.1789   &33.1884    &51.6100\\
15  &26.7008   &35.0946    &52.5281\\
16  &28.2244   &37.0844    &56.2657\\
17  &29.7513   &39.1556    &60.3598\\
18  &31.2706   &41.0815    &62.7705\\
19  &32.7936   &43.0182    &64.7816\\
\hline \hline
\label{tab:mode}
\end{tabular*}
\end{table}

%%%%%%%%%%%%%%%%%%%%%%%%%%%%%%%%%
\section{Results and Discussions}
\label{RD}
%%%%%%%%%%%%%%%%%%%%%%%%%%%%%%%%%
The radial profile for the $\eta$ and $\zeta$ are plotted in Fig. \ref{fig:eta_zeta} at maximum corresponding masses [i.e., at different central densities ($\rho_c$)] for three EOSs with $k_f^{\rm DM}=0.00, 0.03$ and 0.05 GeV. Here we represent the behavior of $f$-mode ($n=0$) and 9 excited $p$-modes ($n=1-9$), and the color bar represents the order number $n$ corresponding to different modes. For both $\eta$ and $\zeta$ in the region $0<r<R$, we are getting exactly $n$ nodes for $n$th mode following the Sturm-Liouville system. The oscillation for $\eta$ is directly proportional to the Lagrangian pressure variation $\Delta P$, and therefore a decaying amplitude is observed when it approaches the stellar surface, following Eq. (\ref{eq:delbcP}). Considering the case of $\eta$, the nodes for higher $p$-modes shifts toward the centre when we increase $k_f^{\rm DM}$. The system tends to oscillate consistently in a somewhat stable region close to the equilibrium point \cite{Clemente_2020} because $\eta$ and $\Delta P$ are both continuous. Considering the profile of $\zeta$, all the modes start from zero as it is associated with the radial displacement function and satisfies the boundary condition at the center, Eq. (\ref{eq:delbcr}), and near the surface, the growing amplitude is observed with rapid sign change \cite{sen2022radial}. For DMANS, we observed that the impact is similar. However, the amplitude of higher $p$-modes is significantly reduced compared with no DM, and the positions of the nodes are changed considerably closer to the surface.

Altering $k_f^{\rm DM}$ with IOPB-I EOS, we examine the dependency of eigenfrequencies with central energy densities ($\rho_c$) for the first 20 radial modes in Fig. \ref{fig:freq} and also enumerated their value in Table \ref{tab:mode}. The stability limit is reached when density rises, irrespective of EOS, as shown in the same figure. The star is approaching its maximum mass ($M_{\rm max}$) at the instability point, which is indicated by the existence of a zero eigenvalue for the $f$-mode \cite{Kokkotas_2001}. The critical central energy density for the $k_f^{\rm DM}$ = 0.00 GeV is $2.0022 \times 10^{15}$ g/cm$^3$. However, when we increase the DM fraction, $\rho_{\rm crit}$ increases as EOS becomes softer and oscillates with higher frequencies. For the $k_f^{\rm DM}$ = 0.03 GeV, 0.05 GeV, the $\rho_{\rm crit}$ = $2.2022 \times 10^{15}$ g/cm$^3$ and $2.9177 \times 10^{15}$ g/cm$^3$ respectively.

At lower densities, $\gamma$ being constant, NS acts as a homogeneous, nonrelativistic body. The angular frequency of oscillation follow $\omega^2 \propto\rho(4\gamma-3)$  \cite{Yong_2022,Shapiro_1983, Arnett_1977}. This explains the dip in frequency initially when the star's central density is low enough, as shown in Fig. \ref{fig:freq}.
%%%%%%%%%%%%%%
\begin{figure}
    \centering
    \includegraphics[scale=0.45]{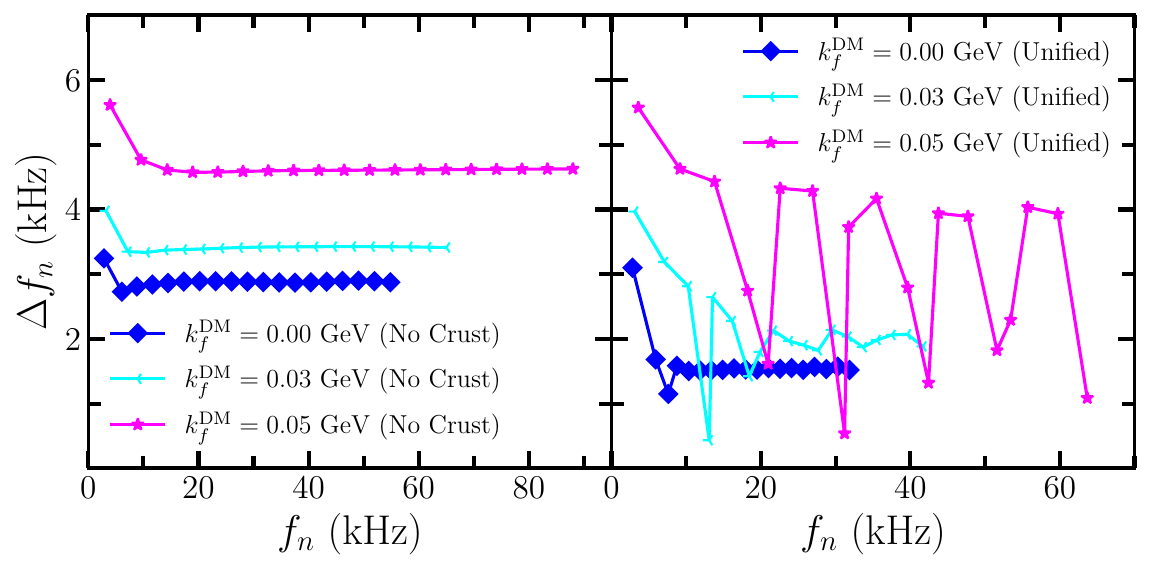}
    \caption{{\it Left:}the frequency differences between consecutive modes for three DM EOSs without crust operating at $1.4 M_\odot$ is made. {\it Right:} the same analysis is done for unified EOSs.}
    \label{fig:freq_diff}
\end{figure}
%%%%%%%%%%%%%%
%%%%%%%%%%%%%%
\begin{figure}
    \centering
    \includegraphics[scale=0.6]{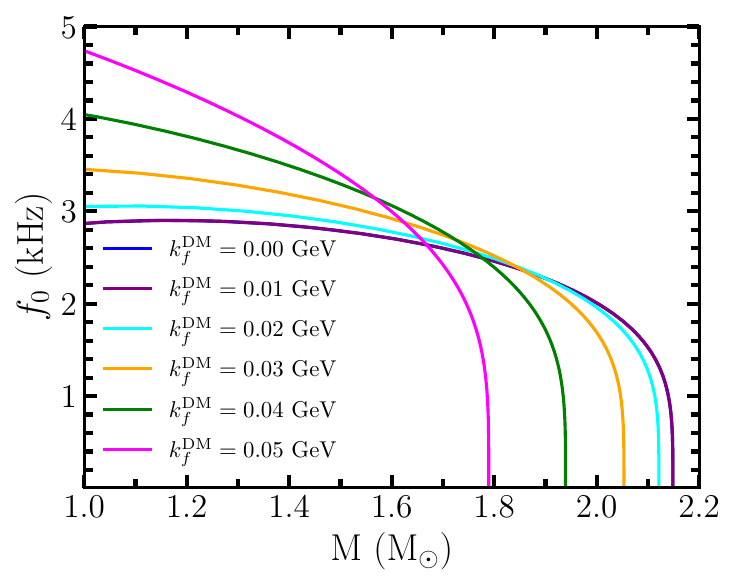}
    \caption{$f$-mode frequency as a function of mass for various DM momentum.}
    \label{fig:mf}
\end{figure}
%%%%%%%%%%%%

Another distinguished observation in Fig. \ref{fig:freq} is that when we increase the DM Fermi momentum, the higher modes oscillation shows various kinks. This depicts an important observation that the frequencies of two successive modes from distinct families rejecting one another as they approach each other, resulting in a sequence of ``avoided crossings" between the respective modes \cite{Kokkotas_2001, Gondek_1997, Gondek_1999}. The eigenvalue problem's solution changes from a standing wave localized mainly in the crust to one primarily localized in the core at the  ``avoided crossings" point\cite{Gondek_1999}. We observed that the phenomenon of avoided crossings is present in all three cases but is much more prominent for larger $k_f^{\rm DM}$. DM makes the NS much more compact leading to a thinner crust, see Fig. \ref{fig:freq}. This allows avoided crossings to happen at lower central density.

In Fig. \ref{fig:freq_diff}, a comparison is made between the frequency difference of two consecutive modes, i.e., $\Delta f_n = f_{n+1}-f_n$ and frequency $f_n$ which is calculated at 1.4 M$_\odot$. Here we take both EOSs with and without crust-varying DM Fermi momentum for better analysis. In the left panel of the figure, we take EOSs without the crust, showing the smooth trend in $\Delta f_n$  and consistent with \cite{Sagun_2020, PhysRevD.98.083001}. Another observation is that for the higher values of $k_f^{\rm DM}$, the magnitude of $\Delta f_n$ is higher. This is because DMANS oscillates with a higher frequency and its magnitude increases with higher percentage of DM inside the star, as seen in Fig. \ref{fig:freq}. However, for unified EOSs (in the right side figure), there is uneven fluctuation in $\Delta f_n$. This is due to the nuclear pasta, which is present inside the inner crust in which the characteristics of the adiabatic index seem to be no more monotonic. Since the crust typically makes up less than $10\%$ of the stellar radius and the oscillation nodes are located deep within the NS core, the radial oscillation lowest order mode ($n=0$) does not get significantly affected by the crust, as seen in Fig. \ref{fig:eta_zeta}. But a few oscillation nodes for higher-order modes are located in the crust. Therefore, the crust considerably modifies the eigenfrequencies, and a peak is displayed by $\Delta f_n$ whenever a node passes through the pasta zone \cite{sen2022radial}.

The variation of $f$-mode frequencies with masses for 6 DM EOSs is shown in Fig. \ref{fig:mf} by varying $k_f^{\rm DM}$. This work directs our attention to a detailed and in-depth investigation of the relationship between radial oscillation and the stability of NS. When we raise the $k_f^{\rm DM}$, as was previously covered in Fig. \ref{fig:freq}, the $f$-mode oscillates here at higher frequencies, which can also be seen in this figure. It is clear that the $f$-mode rapidly approaches zero precisely at the point where maximum numerical NS mass is reached following the $M-R$ profile shown in Fig. \ref{fig:mr} \cite{Sun_2021}. Therefore, the outcome is in line with the stability criteria $\partial M/\partial\rho_c > 0$ \cite{Shapiro_1983, Harrison_1965}. Also, the radial oscillation equations indirectly assist us in demonstrating that increasing the DM Fermi momentum softens the EOS.
%%%%%%%%%%%%%%
\begin{figure}
    \centering
    \includegraphics[scale=0.6]{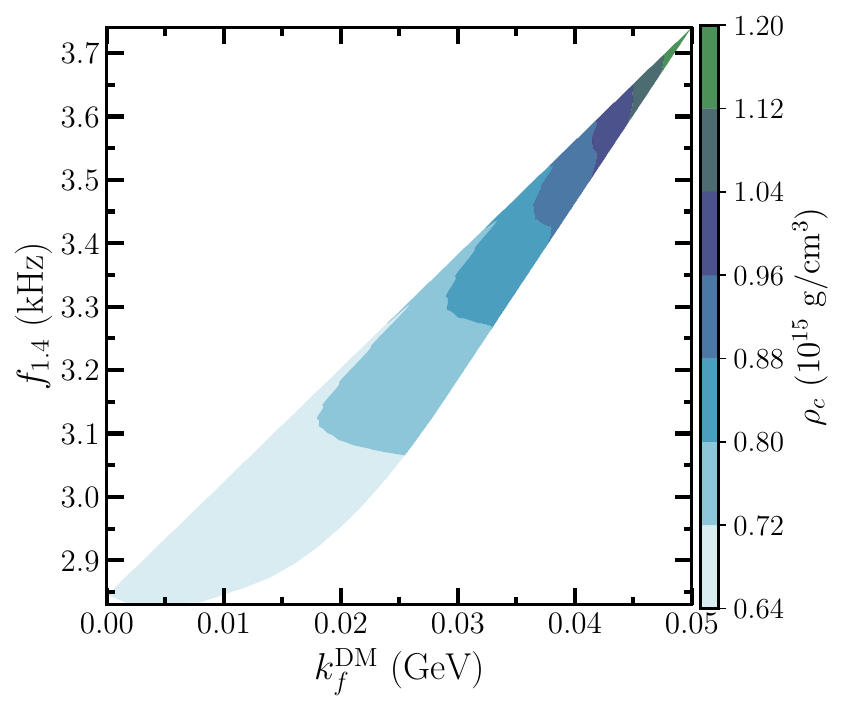}
    \caption{Variation of the canonical $f$-mode frequencies with DM Fermi momenta. The color bar represents the central densities corresponding to $k_f^{\rm DM}$.}
    \label{fig:fmdm_cano}
\end{figure}
%%%%%%%%%%%%

The variation of canonical $f$-mode frequencies with DM Fermi momenta is shown in Fig. \ref{fig:fmdm_cano}. With increasing $k_f^{\rm DM}$ up to 0.025 GeV, there is a slight variation of $f$-mode frequencies. This is because the effects of DM soften the EOSs lesser in magnitude compared to a higher Fermi momentum of DM (see Fig. \ref{fig:EOS}), which slightly decreases the magnitude of the mass. However, the magnitude of $f$-mode frequencies increases more for $k_f^{\rm DM}>0.025$ GeV. Overall, the $f$-mode frequency of canonical NS increases linearly with $k_f^{\rm DM}$. Different color contours show the variation of central densities with $k_f^{\rm DM}$. The careful inspections illustrated that the color bands shrink with increasing $k_f^{\rm DM}$. Also, the area of $\rho_c$ at lower $k_f^{\rm DM}$ is found to be larger as compared to the higher $k_f^{\rm DM}$ and the area slightly reduces from the lower $k_f^{\rm DM}$ to the higher one. This is because the slight increase in $k_f^{\rm DM}$ enhances the value $\rho_c$ a little bit.
When we increase $k_f^{\rm DM}$ by more than 0.025 GeV, the effects of DM on central densities are significant, which decreases the mass very profoundly. For the canonical star, the mass is fixed, but the central densities are different for various $k_f^{\rm DM}$.  Moreover, one can also see the mass variation with $k_f^{\rm DM}$ for other modes, such as $p_1-p_{19}$ modes. However, their variations follow the same trends, but the magnitudes are higher for higher nodes than the $f$-mode case. 
%%%%%%%%%%%%%%%%%%%%%%%%%%%%%%%%%
\section{Summary and Conclusions}
\label{summary}
%%%%%%%%%%%%%%%%%%%%%%%%%%%%%%%%%
To summarize our work, in the present article, we have studied the impact of dark matter on the radial oscillations of nonrotating neutron stars. Assuming fermionic DM matter within supersymmetric models, we have adopted the IOPB-I hadronic EOS. For fixed DM mass and couplings to the nucleons and to the SM Higgs boson, the only free parameter is the DM Fermi momentum, which determines the number density of the DM particles inside the star. First, to describe hydrostatic equilibrium, we solved the structure equations numerically to obtain the mass-radius relationships. Next, to study radial oscillations of pulsating stars, we solved the Sturm–Liouville equations for the perturbations imposing the appropriate boundary conditions, thanks to which we were able to compute the frequencies of the modes as well as the corresponding wave functions. The fundamental $f$-mode and 19 excited $p$-modes have been calculated, with and without DM, varying the DM Fermi momentum. Our numerical results show that the presence of DM inside NSs softens the EOS, and consequently, the maximum mass of the stars is lowered. What is more, adding DM increases the frequencies of pulsating objects, irrespective of the presence of the crust. Finally, the higher the DM Fermi momentum (or, equivalently, the DM mass fraction), the higher the frequencies of the radial oscillation modes.

We also investigated the profile of the eigenfunctions, $\eta$, and $\zeta$, with and without DM, and we found that they oscillate with exactly $n$ nodes for the $n{\rm th}$ mode for both cases. But the presence of DM somewhat affects the position of the nodes. Regarding $\eta$, when the DM varies with higher momenta, the nodes for higher $p$-modes are shifted toward the center, while regarding $\zeta$, the amplitude of higher $p$-modes is significantly reduced when we increase the DM Fermi momentum, and the nodes are relocated considerably closer to the surface. Further, large frequency separation between consecutive modes was studied, varying DM Fermi wave numbers with and without crust, and the effects of crust were noted. Finally, the NS stability was studied varying $f$-mode with mass, and the stability criterion $\partial M/\partial\rho_c > 0$ was verified.

%%%%%%%%%%%%%%%%%%%%%%%
\begin{acknowledgments}
%%%%%%%%%%%%%%%%%%%%%%%
B.K. acknowledges partial support from the Department of Science and Technology, Government of India, with grant no. CRG/2021/000101. T.Z is supported by the Department of Energy, Grant No. DE-FG02-93ER40756.
\end{acknowledgments}
%%%%%%%%%%%%%%%%%%%%%
%%%%%%%%%%%%%%%%%%%%%%%%%
\bibliography{radial.bib}
%%%%%%%%%%%%%%%%%%%%%%%%%
\end{document}